\documentclass[12pt]{article}
\usepackage[dvips]{graphicx}
\usepackage{textcomp}
\usepackage{amssymb}

\newcommand{\mc}{\multicolumn}

\begin{document}
\begin{titlepage}
\vskip0.5cm

\begin{center}
{\Large\bf  Yet an other method to compute the 
thermodynamic Casimir force in lattice models}
\end{center}

\centerline{
\large Martin Hasenbusch
}
\vskip 0.3cm
\centerline{\sl  Institut f\"ur Physik, Humboldt-Universit\"at zu Berlin}
\centerline{\sl Newtonstr. 15, 12489 Berlin, Germany}
\centerline{\sl
e--mail: \hskip 1cm
 Martin.Hasenbusch@physik.hu-berlin.de}
\vskip 0.4cm

\begin{abstract}
We discuss a method that allows to compute the thermodynamic Casimir force
at a given temperature in lattice models by performing a single Monte Carlo 
simulation. It is analogous to the one used by de Forcrand and Noth and
de Forcrand, Lucini and Vettorazzo  in the study of 't Hooft loops and 
the interface tension in SU(N) lattice gauge models in four dimensions.
We test the method at the example of thin films in the XY universality
class. In particular we simulate the improved two-component $\phi^4$ model
on the simple cubic lattice. This allows us to compare with our previous 
study, where we have computed the Casimir force by numerically integrating 
energy densities over the inverse temperature.

\end{abstract}
\end{titlepage}

\section{Introduction}
In 1978  Fisher and de Gennes \cite{FiGe78} realized that  when thermal
fluctuations are restricted by a container a force acts on the walls
of the container. Since this effect is similar to the Casimir effect,
where the restriction of quantum fluctuations induces a force, it is called
``thermodynamic'' Casimir effect. Since thermal fluctuations only extend to
large scales in the neighbourhood of a continuous phase transitions it is
also called ``critical'' Casimir effect. Recently this effect has attracted
much attention, since it could be verified for various experimental systems
and quantitative predictions could be obtained from Monte Carlo
simulations of spin models \cite{Ga09}. The neighbourhood of the critical 
point implies that the Casimir force is described by a universal finite size 
scaling function
\footnote{For  reviews on finite size scaling see \cite{Barber,Privman};
For reviews on critical phenomena and the Renormalization group see e.g. 
\cite{WiKo,Fisher74,Fisher98,PeVi02}.}.
For the film geometry, finite size scaling predicts
\begin{equation}
\label{FSS}
F_{casimir}(L_0,t) \simeq \frac{k_B T}{ L_0^{3}} \theta(t [L_0/\xi_0]^{1/\nu})
\end{equation}
where $k_B$ is the Boltzmann constant, $T$ the temperature, 
$t=(T-T_c)/T_c$ the reduced temperature and
$L_0$ the thickness of the film. The amplitude $\xi_0$ of the correlation
length in the high temperature phase is defined by
\begin{equation}
\xi \simeq \xi_0 t^{-\nu}
\end{equation}
where $\xi$ is the correlation length of the bulk system and $\nu$ its
critical exponent. The function $\theta(x)$ is the same for all films
in a given universality class, where also the boundary universality 
class \cite{Diehl86} has to be taken into account.

As a first application of the numerical method  discussed
here, we study the improved two-component $\phi^4$ model on the simple 
cubic lattice. The phase transition of this model belongs to
the XY universality class in three dimensions.  Also the
$\lambda$-transition of $^4$He shares this universality class.
The experimental study of the $\lambda$-transition provided highly 
accurate estimates for critical exponents and amplitude ratios of the 
bulk system. For a review see \cite{BaHaLiDu07}. Also confined systems 
have been studied in detail at the $\lambda$-transition of $^4$He 
\cite{GaKiMoDi08}. In particular, the thermodynamic Casimir force in thin films 
of $^4$He has been measured \cite{GaCh99,GaScGaCh06}.  
These experiments confirm that the thermodynamic
Casimir force for films of different thickness $L_0$ can indeed 
be described by the same scaling function $\theta(x)$.  For all temperatures
the force turns out to be negative. In the high temperature phase 
$\theta(x)$ is monotonically decreasing with decreasing $x$.
The Casimir force vanishes for large values of $x$. At the 
critical point of the bulk system $\theta(0) = -0.07\pm 0.03$ \cite{GaCh99}. 
In the low 
temperature phase the finite size scaling function shows a minimum 
at $x_{min} \approx -5.5$ with $\theta_{min} \approx 1.3$ \cite{GaScGaCh06}. 
For $x<x_{min}$ the finite size scaling function increases with decreasing 
temperature. For small values of $x$ it seems to approach a finite negative
value.

It has been a challenge for theorists to compute the finite size scaling
function $\theta(x)$. Krech and Dietrich \cite{KrDi92,KrDi92b} have computed it
in the high temperature phase using the $\epsilon$-expansion up to 
O($\epsilon$). This result is indeed consistent with the measurements on 
$^4$He films. Deep in the low temperature phase, the spin wave approximation
should provide an exact result. It predicts a negative non-vanishing value for 
$\theta(x)$. However the experiments suggest a much  larger absolute value
for $\theta(x)$ in this region.  Until recently a reliable theoretical 
prediction for the minimum of $\theta(x)$ and its neighbourhood was missing.
Using a renormalized mean-field approach the authors of \cite{Kardar,MaGaBi07}
have computed $\theta(x)$ for the whole temperature range. Qualitatively
they reproduce the features of the experimental result. However
the position of the minimum is by almost a factor of 2 different from
the experimental one. The value at the minimum is wrongly estimated by a
factor of about 5.

Only quite recently Monte Carlo simulations of the
XY model on the simple cubic lattice \cite{VaGaMaDi07,Hu07,VaGaMaDi08}
provided results for $\theta(x)$ which essentially reproduce the experiments
on $^4$He films \cite{GaCh99,GaScGaCh06}.  In \cite{myCasimir} we have 
applied the method used by \cite{Hu07} to study the improved 
two-component $\phi^4$ model on the simple cubic lattice. The study of this 
model should provide more accurate results since corrections 
$\propto L_0^{-\omega}$ with $\omega =0.785(20)$ \cite{recentXY} are 
eliminated.
Essentially our result confirms those of \cite{VaGaMaDi07,Hu07,VaGaMaDi08}. 
However there is a discrepancy in the position $x_{min}$ 
of the minimum of $\theta(x)$ that is clearly larger than the 
errors that are quoted: In \cite{Hu07} $x_{min} = -5.3(1)$ and in
\cite{VaGaMaDi08} $x_{min} = -5.43(2)$ which has to be compared with
our result $x_{min}=- 4.95(3)$ \cite{myCasimir}.  

In order to verify our result \cite{myCasimir} we compute the thermodynamic 
Casimir force in the two-component $\phi^4$ model using a different method
that is analog to that of \cite{Forcrand1,Forcrand2} used to compute the 
string tension and 't Hooft loops in lattice gauge model. The general idea 
is similar to that of \cite{VaGaMaDi07,VaGaMaDi08}. However, 
in contrast to \cite{VaGaMaDi07,VaGaMaDi08} a single simulation is 
sufficient \footnote{Provided that $f_{bulk}$ is known.} 
to obtain the Casimir force at a given temperature.

This paper is organized as follows: First we define the $\phi^4$ model on 
the simple cubic lattice. Then we discuss in detail the method used here
to compute the thermodynamic Casimir force.
In section \ref{numerical} we discuss our numerical simulations.
First we performed numerical simulations at the critical point 
of the three-dimensional system. Next we computed the free energy density 
for the thermodynamic limit of the three-dimensional system 
at two values of the inverse temperature $\beta$ in the high and the 
low temperature phase each. Then we have measured the thermodynamic 
Casimir force for $L_0=8.5$ at various temperatures. Finally we 
have simulated at $x_{min}$ for the thicknesses $L_0=6.5$, $7.5$, $9.5$, 
$12.5$ and $24.5$ to complement our results of ref. \cite{myCasimir}.
Finally we summarize our results and give our conclusion.

\section{The model and the observables}
\label{phi4model}
We study the two-component $\phi^4$ model on the simple cubic lattice.
We  label the sites of the lattice by
$x=(x_0,x_1,x_2)$. The components of $x$ might assume the values
$x_i \in \{1,2,\ldots,L_i\}$.  We simulate lattices
of the size $L_1=L_2=L$ and $L_0 \ll L$.  
In 1 and 2-direction periodic boundary conditions are used.
In order to mimic the vanishing order parameter that is observed at
the boundaries of $^4$He films, free boundary conditions
in 0-direction are employed. This means that the sites with $x_0=1$ and
$x_0=L_0$ have only five nearest neighbours.
This type of boundary conditions could be interpreted as Dirichlet
boundary conditions with $0$ as value of the field at $x_0=0$ and $x_0=L_0+1$.
Note that viewed this way, the thickness of the film is $L_0+1$ rather
than $L_0$. This provides a natural explanation of the result $L_s=1.02(7)$
obtained in \cite{myKTfilm}.
The Hamiltonian of the two-component $\phi^4$ model, for a vanishing
external field, is given by
\begin{equation}
\label{hamiltonian}
{\cal H} = - \beta \sum_{<x,y>} \vec{\phi}_x \cdot \vec{\phi}_y
 + \sum_{x} \left[\vec{\phi}_x^2 + \lambda (\vec{\phi}_x^2 -1)^2   \right]
\end {equation}
where the field variable $\vec{\phi}_x$ is a vector with two real components.
 $<x,y>$ denotes a pair of nearest neighbour sites on the lattice.
The partition function is given by
\begin{equation}
Z =  \prod_x  \left[\int d \phi_x^{(1)} \,\int d \phi_x^{(2)} \right] \, \exp(-{\cal H}).
\end{equation}
Note that following the conventions of our previous work, e.g. \cite{ourXY},
we have absorbed the inverse temperature $\beta$ into the Hamiltonian.
\footnote{Therefore, following \cite{Fisher98} we actually should call it
reduced Hamiltonian.}
In the limit $\lambda \rightarrow \infty$ the field variables are fixed to
unit length; Hence the XY model is recovered. For $\lambda=0$ we get the exactly
solvable Gaussian model.  For $0< \lambda \le \infty$ the model undergoes
a second order phase transition that belongs to the XY universality class.
Numerically, using Monte Carlo simulations and high-temperature series
expansions, it has been shown that there is a value $\lambda^* > 0$, where
leading corrections to scaling vanish.  Numerical estimates of $\lambda^*$
given in the literature are $\lambda^* = 2.10(6)$ \cite{HaTo99},
$\lambda^* = 2.07(5)$  \cite{ourXY} and most recently $\lambda^* = 2.15(5)$
\cite{recentXY}.  The inverse of the critical temperature $\beta_c$ has been
determined accurately for several values of $\lambda$ using finite size
scaling (FSS) \cite{recentXY}.

We shall perform our simulations at
$\lambda =2.1$, since for this value of $\lambda$ comprehensive Monte
Carlo studies of the three-dimensional system in the low and the
high temperature phase have been performed
\cite{myKTfilm,recentXY,myAPAM,myamplitude}.
At $\lambda =2.1$ one gets $\beta_c=0.5091503(6)$ \cite{recentXY}.
Since  $\lambda =2.1$  is not exactly equal to $\lambda^*$, there are
still corrections $\propto L_0^{-\omega}$, although with a small amplitude.
In fact, following \cite{recentXY}, it should be by at least a factor
20 smaller than for the standard XY model.

\subsection{The energy density and the reduced free energy}
Note that in eq.~(\ref{hamiltonian})  $\beta$ does not
multiply the second term. Therefore, strictly speaking, $\beta$ is not
the inverse of $k_B T$.
In order to study universal quantities it is not crucial how the transition
line in the $\beta$-$\lambda$ plane is crossed, as long as this path is
not tangent to the transition line.
Therefore, following computational convenience, we vary $\beta$ at fixed
$\lambda$. In the following equations it is understood that $\lambda$ 
is kept fixed.

The reduced free energy density is defined as
\begin{equation}
\label{fdef1}
f(\beta) \equiv - \frac{1}{L_0 L_1 L_2} [\log Z(\beta) - \log Z(0)] \;.
\end{equation}
Note that compared with the free energy density $\tilde f$, a factor $k_B T$ 
is skipped.  For convenience we have defined the reduced free energy such
that $f(0)=0$.  For $\beta=0$ the partition function factorizes and 
thus $\log Z(0)/(L_0 L_1 L_2)$ does not depend on the system size.

We define the (internal) energy density as the derivative of
the reduced free energy density with respect to $\beta$.
Furthermore, to be consistent with our previous work e.g. \cite{myheat},
we multiply by $-1$:
\begin{equation}
\label{Edef1}
E = \frac{1}{L_0 L_1 L_2} \frac{\partial \log Z}{\partial \beta} \;.
\end{equation}
It follows
\begin{equation}
\label{Edef}
 E =  \frac{1}{L_0 L_1 L_2}
\left \langle  \sum_{<x,y>} \vec{\phi}_x \cdot \vec{\phi}_y \right \rangle \;,
\end{equation}
which can be easily determined in Monte Carlo simulations.  From
eqs.~(\ref{fdef1},\ref{Edef1}) it follows that the free energy density
can be computed as
\begin{equation}
\label{integrateF}
 f(\beta) = f(\beta_0) - \int_{\beta_0}^{\beta}
              \mbox{d} \tilde \beta   E(\tilde \beta)   \;\;.
\end{equation}

\section{The numerical method}
From a thermodynamic point of view, the Casimir force per unit area is
given by
\begin{equation}
\label{defineF}
F_{casimir} = - k_B T \frac{ \partial f_{ex} }{ \partial L_0} \;\;
\end{equation}
where $L_0$ is the thickness of the film and
$  f_{ex}  =  f_{film} - L_0  f_{bulk}$
is the reduced excess free energy per area of the film.  In lattice models
the thickness $L_0$ assumes only integer values. Therefore we have 
to approximate the derivative by a finite difference 
$F_{casimir}(L_0,t)  \approx - k_B T  \Delta f_{ex}(L_0,t)$,  where
\begin{equation}
\label{finite}
\Delta f_{ex}(L_0,t) \equiv  f(L_0+1/2,t) - f(L_0-1/2,t) - f_{bulk}(t)
\end{equation}
where $L_0+1/2$ is integer.  
$f(L_0+1/2,t)$ and $f(L_0-1/2,t)$ are the reduced free energies of films of the
thicknesses $L_0+1/2$ and $L_0-1/2$, respectively, and $f_{bulk}(t)$ is 
the reduced free energy density of the three-dimensional bulk system. The 
main numerical task is to compute the difference of the reduced free energy 
per area for films of the thickness $L_0-1/2$ and $L_0+1/2$.

In order to compute this difference, it is useful to have the same 
number of field-variables for both systems.  To this end,
we add $L_1 \times L_2$ isolated sites to the film of the thickness 
$L_0-1/2$. Isolated means that the field  $\vec{\phi}$ at such a site is 
subject 
to the potential $\vec{\phi}^2 + \lambda (\vec{\phi}^2 -1)^2$ but the
interaction with other sites is missing. Using our definition~(\ref{fdef1}),  
adding isolated sites to the 
film does not change the free energy per area.  Let us denote the partition 
function of this system by $\bar{Z}_{L_0-1/2}$. Now we can express the 
difference of the reduced free energies as
\begin{eqnarray}
F(L_0+1/2,t) &-& F(L_0-1/2,t) = \log \frac{\bar{Z}_{L_0-1/2}}{Z_{L_0+1/2}} 
\nonumber \\
&=& \log \frac{\mbox{D}[\phi] \exp(-H_{L_0+1/2} ) 
 \exp(-\beta \sum_{<x,y> \in [L_0+1/2]}  \vec{\phi}_x \vec{\phi}_y) }
             {\mbox{D}[\phi] \exp(-H_{L_0+1/2})} \nonumber \\
 &=& \log \; \left \langle 
 \exp(-\beta \sum_{<x,y> \in [L_0+1/2]}  \vec{\phi}_x \vec{\phi}_y) 
     \right  \rangle_{L_0+1/2}  \;\;,
\end{eqnarray}
where $<x,y> \in [L_0+1/2]$ denotes the sum over all nearest neighbour 
pairs, where at least one of the sites is an element of the layer $x_0=L_0+1/2$.
Formally we have rewritten the difference of free energies as an expectation 
value. The problem is that the observable is strongly fluctuating
and therefore it is impossible to obtain an accurate estimate from a 
Monte Carlo simulation of the film of the thickness $L_0+1/2$.
A well known method to overcome this problem is the so called ``multistage
sampling'' strategy; see e.g. \cite{KKMon}.  This means that a sequence 
of systems is introduced that interpolates between the two we are 
interested in. These systems are characterized by the Hamiltonians
$H_0, H_1,\ldots,H_N$, where we identify $H_0=\bar{H}_{L_0-1/2}$ and 
$H_N=H_{L_0+1/2}$. Now we can rewrite the ratio of partition functions as
\begin{equation}
 \frac{Z_0}{Z_N} = \frac{Z_0}{Z_1} \frac{Z_1}{Z_2} \ldots \frac{Z_{N-1}}{Z_N}
 \;\;.
\end{equation}
where we can write the factors as
\begin{equation}
\label{definez}
z_{i+1} \equiv \frac{Z_i}{Z_{i+1}} = \langle \exp(-H_{i+1}+H_i)  \rangle_{i+1} 
\end{equation}
and hence
\begin{equation}
\label{sumf}
F(L_0+1/2,t)-F(L_0-1/2,t) =  \sum_{i=1}^N \log z_i  \;.
\end{equation}
If the sequence is properly chosen and $N$ sufficiently large, the 
fluctuations of $\exp(-H_{i+1}+H_i)$ are small and the expectation value 
can be accurately determined from the simulation of the system 
$i+1$. \footnote{It might be even better to express the difference 
as an expectation value in a system that is between $i$ and $i+1$.}
Obviously there is much freedom in the construction of the sequence of 
systems. A straight forward one is given by
\begin{equation}
 H_i = \bar{H}_{L_0-1/2} + \frac{i}{N} \beta 
 \sum_{<x,y> \in [L_0+1/2]}  \vec{\phi}_x \vec{\phi}_y  \;.
\end{equation}
This choice is very similar to the one used by \cite{VaGaMaDi07,VaGaMaDi08}. 
The main difference is that these authors did consider as starting system 
a film of thickness $L_0-1/2$ plus a two-dimensional system of the 
size $L_1 \times L_2$. This means that in contrast to our choice the 
intra-layer couplings are switched on.

Here we use a different interpolation. It is inspired by a method used to
compute the string tension and 't Hooft loops in lattice gauge theories
\cite{Forcrand1,Forcrand2}.

\begin{figure}
\begin{center}
\scalebox{0.62}
{
\includegraphics{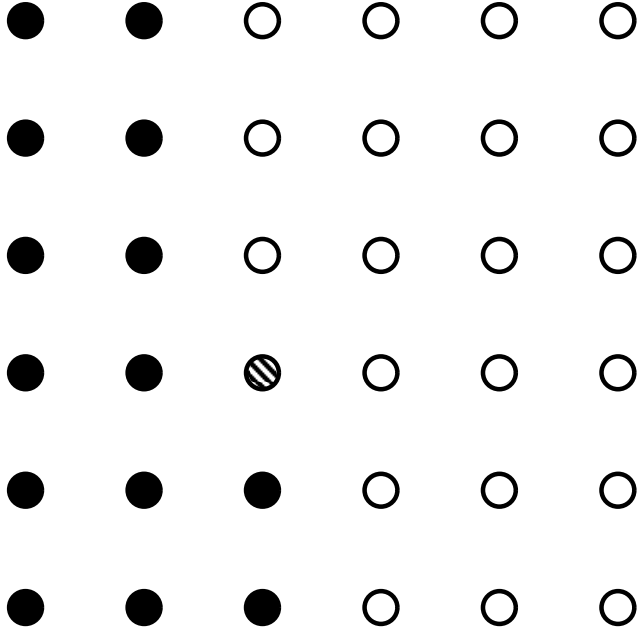}
}
\end{center}
\caption{
\label{lat}  We sketch the layer $x_0=L_0+1/2$ of our system.  In the 
sketch $L_1=L_2=6$.  The sites are given by circles. The filled ones 
are coupled to the system, while the empty ones are isolated.  
We compute the free energy difference between the system, where the 
shaded circle is isolated and the system, where it is coupled to the film.
For a discussion see the text.
}
\end{figure}

We add the isolated sites one by one to the film.  
In the step $i= x_1 L_2 + x_2$ the site $x=(L_0+1/2,x_1,x_2)$ is added. 
All sites that are added are coupled with their nearest neighbours that 
are already in the film. This way we have constructed a sequence of 
$L_1 \times L_2 +1$ systems. Hence,  $L_1 \times L_2$ independent Monte 
Carlo simulations have to be performed to obtain $F(L_0+1/2,t)-F(L_0-1/2,t)$.
Following \cite{Forcrand1,Forcrand2} this can however be avoided:
With increasing $L_1, L_2$ 
the sum~(\ref{sumf}) is dominated by contributions where the newly added
site is far from the boundaries in 1 and 2-direction. 
Hence most of the contributions are essentially
equal to that for $x_1=L_1/2$ and  $x_2=L_2/2$ as 
sketched in figure \ref{lat}. In the limit $L_1, L_2 \rightarrow \infty$, 
this should become exact.   Hence only a single simulation for 
$x_1=L_1/2$ and $x_2=L_2/2$ is required.

Usually updates are performed on the whole lattice
\footnote{In the Monte Carlo slang this is often called ``sweep''.}
before a measurement of the observables is performed.
However in the present case, the observable is localized at a single site.
Therefore the effort for the measurement and the 
update would be highly unbalanced. To circumvent this problem,  
one would like to update the fields in the neighbourhood 
of this site more frequently than those  far off. In order to 
achieve this we follow the idea presented in \cite{ourhierarchy}: 
We consider a sequence of sub-sets of the sites of the lattice.
Here, the smallest set consists of the site $(L_0+1/2,L_1/2,L_2/2)$ only. 
The next 
larger one consists of $(L_0+1/2,L_1/2,L_2/2)$ and its three neighbours 
$(L_0+1/2,L_1/2-1,L_2/2)$,  $(L_0+1/2,L_1/2,L_2/2-1)$ and 
$(L_0-1/2,L_1/2,L_2/2)$. The larger 
ones are given by blocks of the size $b_l \times (2 b_l+1) \times (2 b_l+1)$
and $L_0 \times (2 b_l+1) \times (2 b_l+1)$ if $b_l>L_0$. These blocks
are centred around the site $(L_0+1/2,L_1/2,L_2/2)$. If $b_l<L_0$, 
the eight corners of these 
blocks are $(L_0+1/2,L_1/2-b_l,L_2/2-b_l)$, $(L_0+1/2,L_1/2-b_l,L_2/2+b_l)$, 
$(L_0-1/2,L_1/2-b_l,L_2/2+b_l)$, $(L_0-1/2,L_1/2+b_l,L_2/2+b_l)$, 
$(L_0+1/2-b_l,L_1/2-b_l,L_2/2-b_l)$,    $(L_0+1/2-b_l,L_1/2-b_l,L_2/2+b_l)$,
$(L_0+1/2-b_l,L_1/2-b_l,L_2/2+b_l)$ and $(L_0+1/2-b_l,L_1/2+b_l,L_2/2+b_l)$.
In our simulations we have used $b_l = 1,2,3,5,10,20,40,80, \ldots$, 
where the largest $b_l$ is chosen such that $b_l < L_1,L_2$.  

In a certain sequence, Metropolis and overrelaxation sweeps 
\footnote{We have implemented these updates as discussed in appendix A of
\cite{ourXY}}
over these sub-sets are performed.
This sequence, which we shall call one update cycle, is best explained by 
the following pseudo-code
\begin{verbatim}
cluster_update(); metrosweep(full lattice); oversweep(full lattice);
for(i1=0; i1 < m_1; i1++)
  {
  metrosweep(b_1); oversweep(b_1);
  for(i2=0; i2 < m_2; i2++)
    {
    metrosweep(b_2); oversweep(b_2);
    .
    .
    .
        for(iM=0; iM < m_M; iM++)
          {
          metrosweep(b_M); oversweep(b_M);
          measure();
          }
    .
    .
    .
    }
  }
\end{verbatim}

We did not perform an accurate tuning of the parameters 
$m_1, m_2, m_3,\ldots m_M$; instead we have chosen them such that the 
CPU-time spent at each block-size is roughly the same. 

In the case of the single cluster-updates \cite{wolff} it is easy to focus 
on the
site $(L_0+1/2,L_1/2,L_2/2)$. One simply starts the clusters at the site 
$(L_0+1/2,L_1/2,L_2/2)$ instead of choosing the starting point at random.
In our numerical tests we have not yet implemented this idea.

\subsection{The measurement}
The measurement consists in its most naive implementation in the
evaluation of 
\begin{equation}
A = \exp(-\beta \vec{\phi}_{(L_0+1/2,L_1/2,L_2/2)} \cdot \vec{\Phi}_{(L_0+1/2,L_1/2,L_2/2)})
\end{equation}
where
\begin{equation}
\vec{\Phi}_{(L_0+1/2,L_1/2,L_2/2)} =
  \vec{\phi}_{(L_0+1/2,L_1/2-1,L_2/2)} + \vec{\phi}_{(L_0+1/2,L_1/2,L_2/2-1)}
+ \vec{\phi}_{(L_0-1/2,L_1/2,L_2/2)}
\end{equation}
We have reduced the variance by performing the integral over the 
angle of the field $\vec{\phi}_{(L_0+1/2,L_1/2,L_2/2)}$ 
exactly. This results in the improved observable
\begin{equation}
\label{integra}
\bar{A} = \frac{1}{\int_{0}^{2\pi} \mbox{d} \alpha  \; \exp(-R \cos{\alpha} )}
  =  \frac{1}{2 \pi I_0(R)}
 \end{equation}
 where 
 \begin{equation}
R = \beta |\vec{\phi}_{(L_0+1/2,L_1/2,L_2/2)}| 
          |\vec{\Phi}_{(L_0+1/2,L_1/2,L_2/2)}|
 \end{equation}
and $I_0(R)$ is a modified Bessel function. For our simulations we have 
tabulated $1/(2\pi I_0(R))$  for $0 \le R \le 3$ with a step-size of $0.0001$,
i.e. for $30001$ values of $R$. 
During the simulation we then evaluated $1/(2\pi I_0(R))$ for
$0 \le R \le 3$  by quadratically interpolating the results given in the 
table.
If $R>3$ we have evaluated the integral in eq.~(\ref{integra}) numerically.
A sufficient precision can already be achieved with about 30 nodes.

The expectation value
\begin{equation}
z=\langle A \rangle = \langle \bar{A} \rangle 
\end{equation}
is estimated by averaging $\bar{A}$ over all measurements that we performed 
after thermalization. Here we have dropped the subscript $i=(L_1/2) L_2 +L_2/2$
of eq. \ref{definez}, since only this value of $i$ will be considered in the 
following.   During the simulation we have averaged already all 
measurements in a given update cycle. These averages were written to a file.
The statistical error was then computed taking into account the integrated
autocorrelation time of these cycle averages.

\section{Numerical Results}
\label{numerical}
First we have simulated at the critical temperature of the bulk system.
Next we have determined the reduced free energy of the 
bulk system  at  $\beta=0.49$ and $\beta=0.5$ in the high temperature phase
and at $\beta=0.533$ and $\beta=0.56$ in the low temperature phase.
For $L_0,L_1,L_2 \gg \xi$ the reduced free energy of the bulk system is given 
by $f_{bulk} = \log z$. Our new results are consistent with those obtained 
by integrating the energy densities computed in \cite{myheat}.  Then we have
studied films of the thickness $L_0=8.5$ at four temperatures in the 
low temperature phase of the bulk system. Also here we find that the  
results are consistent with those of \cite{myheat}. Finally we have 
simulated the thicknesses $L_0=6.5$, $7.5$, $9.5$, $12.5$ and 
$24.5$ at $x_{min}$. These simulations complement our results of \cite{myheat}
at $x_{min}$.

As random number generator we have used the  SIMD-oriented Fast
Mersenne Twister algorithm \cite{twister}.

\subsection{Simulations at the critical point}
First we performed simulations at the inverse critical temperature 
$\beta_c=0.5091503(6)$ of the three-dimensional system  using 
lattices of the thicknesses
$L_0=8.5$, $12.5$, $16.5$, $24.5$, $32.5$ and $64.5$.  In all cases we have
chosen $L_1=L_2=12.5 \times (L_0-1/2)$.  Since the correlation length 
of the film is $\xi_{2nd,film}/L_{0,eff} \approx 0.416$ \cite{myheat} 
this should be sufficient to keep deviations from the two-dimensional 
thermodynamic limit smaller than our statistical errors.  As a check we have
simulated for $L_0=8.5$ in addition $L_1=L_2=20,30$ and $50$. We find 
$z=0.84950517(36)$, $0.84951362(36)$ and $0.84951572(37)$ for these lattice
sizes, respectively.  Indeed, starting from $L_1=L_2=30$ our results are
consistent within error bars.
Our results for $L_1=L_2=12.5 \times (L_0-1/2)$ are
summarized in table \ref{betac}.  In these simulations we have used  
block sizes up to $b_1=10$, $20$, $20$, $40$,  $40$ and $80$ for 
$L_0=8.5$, $12.5$, $16.5$, $24.5$, 
$32.5$ and $64.5$, respectively. For all these thicknesses and for all  
block-sizes we have used $m_l=6$. The number of update cycles is
 21349600, 7738500, 7700000, 2040000,  1308200 and  281700 for
$L_0=8.5$, $12.5$, $16.5$, $24.5$, $32.5$ and $64.5$, respectively.
In total these simulations took about 16 month of CPU-time on 
a single core of a Quad-Core Opteron(tm) 2378 CPU (2.4 GHz).

\begin{table}
\caption{\sl \label{betac}
Results for  $z$ at $\beta_c=0.5091503$ \cite{recentXY} for
lattices of the size $L_1=L_2=12.5 \times L_0$.
 }
\begin{center}
\begin{tabular}{|r|r|}
\hline
\multicolumn{1}{|c}{$L_{0}$}&
\multicolumn{1}{|c|}{$z$} \\
\hline
 8.5  &  0.84951552(24) \\
12.5  &  0.84947657(16) \\
16.5  &  0.84946525(23) \\
24.5  &  0.84945897(13) \\
32.5  &  0.84945717(16) \\
64.5  &  0.84945602(14) \\  
\hline
\end{tabular}
\end{center}
\end{table}

The reduced excess free energy behaves as 
\begin{equation}
f_{ex}(L_0,t) = L_{0,eff}^{-2}  h(t [L_0/\xi_0]^{1/\nu}) + b(t) 
\end{equation}
where $L_{0,eff}=L_0+L_s$ with $L_s=1.02(7)$ \cite{myKTfilm} takes
into account corrections due to the Dirichlet boundary conditions and 
$b(t)$ gives the effect of the Dirichlet boundary conditions on the 
analytic part of the free energy of the film. For a discussion and 
references see \cite{myalternative}.
Taking the derivative with respect to $L_0$ at $t=0$ we arrive at
\begin{equation}
- \left . \frac{\partial f_{ex}(L_0,t)}{\partial L_0} \right|_{t=0} = 
 2 h(0) L_{0,eff}^{-3}   = \theta(0)  L_{0,eff}^{-3}
\end{equation}
where $\theta$ is the finite size scaling function of the 
thermodynamic Casimir force.

It follows   
\begin{equation}
\label{betacfit}
\log z(L_0,\beta_c) =  f_{ns} (\beta_c) - \theta(0) L_{0,eff}^{-3}. 
\end{equation}
Note that in the thermodynamic limit the singular part of the 
free energy density vanishes at the critical point; hence 
$f_{bulk}(\beta_c)=f_{ns} (\beta_c)$. The results of our fits 
are given in table \ref{fitbc}.

\begin{table}
\caption{\sl \label{fitbc}
Results for fits with the ansatz~(\ref{betacfit}), where we have used 
$L_s = 1.02$ as input. All data for $L_0 \ge L_{0,min}$ are  fitted
For a discussion see the text.
}
\begin{center}
\begin{tabular}{|r|r|r|c|}
\hline
 \mc{1}{|c}{$L_{0,min}$} & 
 \mc{1}{|c}{$f_{ns}$}   &     
 \mc{1}{|c}{$\theta(0)$}  &   
 \mc{1}{|c|}{$\chi^2/$d.o.f.} \\
\hline
  8.5 &--0.16315935(9)\phantom{0} &--0.0606(3)\phantom{0}   &  0.20 \\
 12.5 &--0.16315932(10)           &--0.0603(6)\phantom{0}   &  0.14 \\
 16.5 &--0.16315930(12)           &--0.0597(17)             &  0.15 \\
\hline
\end{tabular}
\end{center}
\end{table}

In order to estimate the effect of the error of $L_s$ on our results 
we have repeated these fits using $L_s=0.95$. 
E.g. for $L_{0,min}=12.5$
we get $f_{ns}=-0.16315930(10)$ and $\theta(0)=-0.0593(5)$.  
We have also checked 
the effect of the error of $\beta_c$. To this end we have computed 
$\Delta f(L_0,0.5091509)$ by using the data for the energy given in table 1
of \cite{myalternative}. 
We find that the effect on $f_{ns}$ and  $\theta(0)$ is small and can be 
ignored 
here. Based on the result obtained for $L_{0,min}=12.5$ we take as final results
\begin{equation}
 f_{ns}=-0.1631593(1) \;\;\;\; , \;\; \theta(0)=-0.060(2) 
\end{equation}
where the error-bar covers both the statistical error as well as the error
due to the uncertainty of $L_s$. 

This can be compared with the result for $^4$He films  
$\theta(0) = -0.07\pm 0.03$ \cite{GaCh99}, the $\epsilon$-expansion 
up to O($\epsilon$): $\theta(0) = -0.044$ taken from table I of 
\cite{KrDi92} and the estimate $\theta(0)=-0.062(5)$
obtained from Monte Carlo simulations of the standard XY model \cite{Hu07}.  
The authors of \cite{MoNi87} quote $h(0) \simeq -0.03$ (in their notation
$\Delta^f$) as final result.  All these results are consistent with ours.
The largest discrepancy is seen for the $\epsilon$-expansion. However 
one should
note that in \cite{Hu07,myCasimir} it has been observed that in the high 
temperature phase for $x \gtrapprox 1$ the numerical 
result for $\theta$ matches nicely with the $\epsilon$-expansion \cite{KrDi92}.

\subsection{Free energy density of the bulk system}
\label{bulkf}
Here we compute the free energy density of the bulk system for 
two values of $\beta$ in the high temperature phase and two 
values of $\beta$ in the low temperature phase. These results are 
compared with ones obtained by integrating the energy density starting 
from $\beta_c=0.5091503$ using the start value $f(\beta_c)=-0.1631593(1)$ 
obtained above.

For sufficiently large $L_0$, $L_1$ and $L_2$  the quantity $\log z$ 
should be a good 
approximation of the bulk free energy density.  In particular in the 
high temperature phase, this should be the case for $L_0,L_1,L_2 \gg \xi_{3D}$.
Here we performed simulations at $\beta=0.49$ where 
$\xi_{2nd,3D} =3.72370(19)$  and $\beta=0.5$ where $\xi_{2nd,3D} = 6.1498(5)$
(see table 5 of \cite{myamplitude}).

At $\beta=0.49$ we have simulated $L_0=49.5$, $L_1=L_2=50$  and 
$L_0=99.5$, $L_1=L_2=100$. 
For $L_0=49.5$, $L_1=L_2=50$ we have used block sizes up to $b_1=20$
and $m_l=6$. From  5203000 cycles we get 
$f(0.49) = -0.14712079(18)$.
For $L_0=99.5$, $L_1=L_2=100$ we have used  block sizes
up to $b_1=40$ and $m_l=6$. From  945000 cycles we get
$f(0.49) = -0.14712095(17)$. 
As expected, 
these results are indeed consistent within error bars and hence a good 
approximation of the thermodynamic limit. 

Based on the experience gained at $\beta=0.49$ we have simulated at 
$\beta=0.5$ only the lattice size $L_0=99.5$, $L_1=L_2=100$. We have 
used  block sizes up to $b_1=20$ and  $m_l=6$. 
From  598000 cycles we get 
$f(0.5) = -0.15519942(24)$. 

In the low temperature phase we find from simulations of a 
$199.5 \times 500^2$ lattice  $f(0.533)= -0.18931867(66)$ and 
$f(0.56)= -0.22693625(73)$. We have used blocks up to the size $b_1=80$
and $m_l=6$ for all block sizes. We performed 24700 and 21700 cycles for 
$\beta=0.533$ and $\beta=0.56$, respectively.
Both of these simulations took about 8 weeks of CPU-time on
a single core of a Quad-Core Opteron(tm) 2378 CPU (2.4 GHz).

Now we can check whether these results for the free energy density are 
consistent with those obtained from integrating 
the energy density \cite{myheat} using eq.~(\ref{integrateF}).

In \cite{myheat} we have computed the energy density of the three-dimensional 
bulk system in the range of inverse temperatures $0.49 \le \beta \le 0.58$.
We have fitted these data  in the range $0.49 \le \beta \le 0.529$ with the 
ansatz
\begin{equation}
\label{criticalE}
 E(\beta) = E_{ns} + C_{ns} (\beta-\beta_c)
       + a_{\pm} |\beta-\beta_c|^{1-\alpha}
       + d_{ns} (\beta-\beta_c)^2
     + b_{\pm} |\beta-\beta_c|^{2-\alpha}
\end{equation}
where $E_{ns}$, $C_{ns}$, $\beta_c=0.5091503(6)$ and
$\alpha=-0.0151(3)$ \cite{recentXY} are input and
$a_{\pm}$, $d_{ns}$ and $b_{\pm}$ are the 5 free parameters of the fit.
For $\beta < 0.529$ 
we have integrated this ansatz, using the results for the fit-parameter
obtained in \cite{myheat}. In all cases we have taken 
$\beta_0=\beta_c=0.5091503$ as starting point of the integration, where 
we have used the estimate of $f(\beta_c)$ obtained above.
Our results are summarized in table \ref{integrateF}. 
For $\beta>0.529$ we performed a numerical integration of the energy density
using the trapezoidal rule, starting from $\beta_0=0.52$. The estimate 
for $f(0.52)$ is taken from table \ref{integrateF}. We have checked 
that our result virtually does not depend on the choice of $\beta_0$, where 
we switch from the integration of the ansatz~(\ref{criticalE}) to the 
numerical integration of the energy density. Also the results for 
$\beta>0.529$  are given in table \ref{integrateF}.
The error quoted is dominated by the error for the free energy at $\beta_c$.

In table \ref{integrateF} we also give our results for the free energy 
density of the bulk system at $\beta=0.49$, $0.50$, $0.533$ and $0.56$
as computed by the new method discussed here. We find that the results
are consistent within error-bars. This confirms that we can indeed 
compute the free energy density of the bulk system with 6 to 7 accurate 
digits.

\begin{table}
\caption{\sl \label{integrateF} Numerical results for the free energy
density of the bulk system. These where obtained by integration of the 
energy density. As starting point of the integration we have taken 
the critical point $\beta_c$ and the value $f(\beta_c)$ obtained in the 
previous subsection. In addition in the third column we give estimates
of the free energy density obtained directly with the method discussed 
in the present work.
}
\begin{center}
\begin{tabular}{|l|l|l|}
\hline
 \mc{1}{|c}{$\beta$}  &   \mc{1}{|c|}{$f$ INTEGRAL} & \mc{1}{|c|}{$f$ DIRECT}     \\
\hline
0.49    &  --0.1471210(1) & --0.1471210(2) \\
0.50    &  --0.1551994(1) & --0.1551994(2) \\
0.51463 &  --0.1684460(1) & \\
0.52    &  --0.1740847(1) & \\
0.52348 &  --0.1779533(1) & \\
\hline
0.5301  &  --0.1857405(1) &  \\  
0.533   &  --0.1893183(1) & --0.1893187(7) \\  
0.53814 &  --0.1958961(2) &  \\ 
0.54    &  --0.1983478(2) &  \\ 
0.54432 &  --0.2041851(2) &  \\
0.56    &  --0.2269362(2) & --0.2269363(7)\\  
\hline
\end{tabular}
\end{center}
\end{table}

\subsection{Films of the thickness $L_0=8.5$}
We have simulated at  $\beta=0.52,0.533,0.54,0.56$ in the low temperature
phase of the three-dimensional system.  We have taken lattices of the size
$L_1=L_2=50$, $100$, $250$, $500$ and $1000$ to control corrections to the 
two-dimensional thermodynamic limit of the thin film.
The simulations for $L=1000$ took about 18 days  of CPU time each.
In table \ref{table85} we give our results for $-\Delta f_{ex}$, where 
we have used the estimates of $f_{bulk}$ computed in section \ref{bulkf}.  
At $\beta=0.52$ the results for all choices of 
$L_1=L_2$ are consistent within error-bars. At $\beta=0.533$ a clear deviation
of the results from those obtained for larger lattices can be observed
up to $L_1=L_2=100$. The result for $L=500$ deviates by a bit more
than two standard deviations from that for $L=1000$, while the results for
$L=250$ and $L=1000$ are consistent within error bars.  For $\beta=0.54$
and $\beta=0.56$ the results obtained for $L=250$, $500$ and $1000$
are consistent within error-bars. We conclude that in all cases 
for $L=1000$ the deviation from the thermodynamic limit is smaller than 
the error bar.

\begin{table}
\caption{\sl \label{table85} Numerical results for minus the  derivative of 
the excess free energy $-\Delta f_{ex}$ of films of the thickness $L_0=8.5$.
In the last row we give the results of \cite{myCasimir} for comparison.
}
\begin{center}
\begin{tabular}{|r|l|l|l|l|}
\hline
 \mc{1}{|c}{$L$ $\backslash$  $\beta$} & \mc{1}{|c}{0.52}   &  \mc{1}{|c}{0.533}
&        \mc{1}{|c}{0.54}          &  \mc{1}{|c|}{0.56}  \\
\hline
  50 & --0.0007423(14) & --0.0014878(23) &  --0.0010417(27) &--0.0003621(23) \\
 100 & --0.0007432(8)  & --0.0015797(13) &  --0.0011867(18) &--0.0003870(13) \\
 250 & --0.0007436(5)  & --0.0015845(8)  &  --0.0012564(14) &--0.0003934(8) \\
 500 & --0.0007439(3)  & --0.0015863(5)  &  --0.0012679(10) &--0.0003940(5) \\
1000 & --0.0007433(3)  & --0.0015846(5)  &  --0.0012666(11) &--0.0003945(6) \\
 \hline
ref. \cite{myCasimir}  & --0.0007392(18) & --0.0015795(24) &  --0.0012600(26) &--0.0003874(28)\\
\hline
\end{tabular}
\end{center}
\end{table}

For comparison we give in the last row results \cite{myCasimir} which
were obtained by numerical integration of Monte Carlo data for $\Delta E_{ex}$.
We see that the results of \cite{myCasimir} are about $0.000004$ larger 
than our present ones. This deviation is about twice the statistical error.
In \cite{myCasimir} we have 
started the integration at $\beta=0.49$ for $L_0=8.5$, setting 
$\Delta f_{ex}(0.49)=0$.  
From the $\epsilon$-expansion \cite{KrDi92}
we get $\theta(x) \approx -0.0039$ for  $x=t [L_{0,eff}/\xi_0]^{1/\nu}$
corresponding to $\beta=0.49$ and $L_0=8.5$. Hence 
$-\Delta f_{ex} =\theta L_{0,eff}^{-3} \approx - 0.0000041$ which fully explains
the difference observed in table \ref{table85}.

\subsection{The minimum of the Casimir force}

In \cite{myCasimir} we have determined the position of the minimum 
of $\theta$ for a large number of thicknesses of the film. To this end
we have determined the zero of
\begin{equation}
\label{deltaE}
\Delta E_{ex}(L_0,\beta) = E(L_0+1/2,\beta) - E(L_0-1/2,\beta)
- E_{bulk}(\beta)
\end{equation}
where $E(L_0+1/2,\beta)$ is the energy per area of a film of the thickness $L_0+1/2$ and
$E_{bulk}(\beta)$ the  energy density of the three-dimensional bulk system.
We had simulated at a few values of $\beta$ in the neighbourhood of 
$\beta_{min}$. To get a preliminary estimate of $\beta_{min}$ we used the 
information gained already from the simulations for $L_0=8.5$, $16.5$ and 
$32.5$, where we have simulated a large range of $\beta$-values and
the ansatz $\beta_{min}(L_0) - \beta_c \simeq L_{0,eff}^{-1/\nu}$.
These results are given in 
table \ref{minimum}, which we have copied from table 2 of \cite{myCasimir}.  
In the present work, 
we have added the values of $\theta_{min}$ for $L_0=6.5$, $7.5$, 
$9.5$, $12.5$ and $24.5$ that were missing in \cite{myCasimir}. To this end,
we have simulated lattices of the size $L_1=L_2=500$ for 
$L_0=6.5$ and $7.5$, $L_1=L_2=1000$ for $L_0=9.5$ and $L_0=12.5$, 
and $L_1=L_2=2000$ for $L_0=24.5$. From these simulation we get
$\log z$, while $f_{bulk}$ is taken from table \ref{integrateF}. 
Our results for $-\Delta f_{ex}$ are given in table \ref{minimum}.

Let us briefly discuss the simulation of the $L=24.5$ film: The simulations
took about 2 month of CPU-time on a single core  of a 
Quad-Core Opteron(tm) 2378 CPU (2.4 GHz).  We performed 33000 update 
cycles.  We have used  block sizes up to  $b_1=160$ and $m_l=6$ for
all block sizes.

\begin{table}
\caption{\sl \label{minimum} 
The position $\beta_{min}$ of the
minimum of the Casimir force and its value $-\Delta f_{ex,min}$ as a
function of the thickness $L_0$.  In the present work we have completed
the table by adding $\Delta f_{ex,min}$  for $L_0=6.5$, $7.5$, $9.5$,
$12.5$, and $24.5$. 
}
 \begin{center}
 \begin{tabular}{|r|c|l|}
 \hline
 \mc{1}{|c}{$L_0$} &
 \mc{1}{|c}{$\beta_{min}$} &
 \mc{1}{|c|}{$- \Delta f_{ex,min}$} \\
 \hline
  6.5 & 0.54432(2) & {\bf --0.0032744(13)} \\
 7.5 & 0.53814(2) &  {\bf --0.0022305(11)}  \\
 8.5 & 0.53354(2) &--0.001582(3) \\
 9.5 & 0.53010(2) &{\bf --0.0011714(8) } \\
 12.5 & 0.52348(2) & {\bf --0.0005468(6)}  \\
 16.5 & 0.51886(2) &--0.0002494(11) \\
24.5 & 0.51463(2) & {\bf --0.0000803(3) } \\
32.5 & 0.51279(2) & --0.0000348(5) \\
\hline
\end{tabular}
\end{center}
\end{table}

First we have fitted the results for $- \Delta f_{ex,min}$ 
given in the third column of table \ref{minimum}  with the ansatz
\begin{equation}
\label{fit1}
 - \Delta f_{ex,min} = \theta_{min}  (L_0 + L_s)^{-3} 
\end{equation}
where $\theta_{min}$ and $L_s$ are the free parameters of the fit.
Our results are summarized in table \ref{fit1}.

\begin{table}
\caption{\sl \label{fitm1} We have fitted the minimum of the thermodynamic 
Casimir force with the ansatz~(\ref{fit1})
    }
\begin{center}
\begin{tabular}{|r|c|l|c|}
\hline 
  $L_{0,min}$    & $\theta_{min}$    & $L_s$ & $\chi^2/$d.o.f. \\
\hline 
    6.5          & --1.299(2)        &   0.849(5)  & 2.64  \\
    7.5          & --1.305(3)        &   0.864(7)  & 1.64  \\
    8.5          & --1.313(5)        &   0.889(13) & 0.89  \\
    9.5          & --1.310(5)        &   0.880(15) & 0.34  \\
   12.5          & --1.312(9)        &   0.888(33) & 0.50 \\
\hline 
\end{tabular}
\end{center}
\end{table}

The $\chi^2/$d.o.f. is smaller than 1 starting from $L_{0,min}=8.5$, where 
all data with $L_0 \ge L_{0,min}$ are included into the fit. We find  
$L_s \approx 0.89$ which is a bit smaller than our previous result 
$L_s = 1.02(7)$  \cite{myKTfilm}. Note that already in \cite{myCasimir} 
we observed that $L_s = 0.95$ apparently leads to a better matching of the 
data than $L_s = 1.02$.

To check the possible effect of sub-leading corrections we have fitted 
our data also with the ansatz
\begin{equation}
\label{fit2}
 - \Delta f_{ex,min} = \theta_{min}  (1+c L_0^{-2}) (L_0 + L_s)^{-3} \;.
\end{equation}
Note that there a number of different corrections with an correction 
exponent close to 2. E.g. $\propto L_0^{-\omega'}$ with $\omega' = 1.8(2)$
\cite{RG} or the restoration of the symmetries that are broken by the 
lattice. Our results are summarized in table \ref{fit2}.

Now the value 
$L_s\approx 0.95$ is fully consistent with our previous result \cite{myKTfilm}.
  As final result we quote $\theta_{min} = -1.31(2)$, where we 
have estimated the systematic error  by the difference of the two 
fits~(\ref{fit1},\ref{fit2}).
This result  fully confirms our previous estimate $\theta_{min} = -1.31(3)$ 
\cite{myCasimir}.

\begin{table}
\caption{\sl \label{fitm1} We have fitted the minimum of the Casimir force
with the ansatz~(\ref{fit2})
  }
   \begin{center}
 \begin{tabular}{|r|c|l|c|c|}
 \hline
 \mc{1}{|c}{$L_{0,min}$}  & 
 \mc{1}{|c}{$\theta_{min}$}    & 
 \mc{1}{|c}{$L_s$}  & 
 \mc{1}{|c}{c} & \mc{1}{|c|}{$\chi^2/$d.o.f.} \\
 \hline
  6.5          & --1.322(8)\phantom{0} & 0.953(3)  & 1.08(35) & 1.13 \\
  7.5          & --1.320(10)& 0.945(5)  & 0.97(61) & 1.40 \\
\hline 
\end{tabular}
\end{center}
\end{table}

\section{Summary and Conclusion}
We have discussed a method to compute the thermodynamic Casimir force in 
lattice models which is
closely related with the one used by de Forcrand and Noth \cite{Forcrand1} 
and de Forcrand, Lucini and Vettorazzo \cite{Forcrand2} in the study of 't Hooft
loops and the interface tension in SU(N) lattice gauge models in four
dimensions.

We have tested the method at the example of thin films of the improved 
two-component $\phi^4$ model on the simple cubic lattice. This model
shares the XY universality class with the $\lambda$-transition of 
$^4$He. Therefore the Casimir force that is measured for thin films
of $^4$He \cite{GaCh99,GaScGaCh06} should be governed by the same 
universal finite size scaling function $\theta$ as that computed from lattice  
models in the XY universality class.

Only quite recently  $\theta$ has been obtained 
from Monte Carlo simulations of the standard XY model on the simple 
cubic lattice \cite{VaGaMaDi07,Hu07,VaGaMaDi08}. This result is of 
particular interest, since other theoretical methods do not  provide
us with accurate results for $\theta$ for the whole range of the 
scaling variable $x=t [L_0/\xi_0]^{1/\nu}$.  Overall one finds a 
reasonable match between the experimental and Monte Carlo results.
In \cite{myCasimir} we have redone the Monte Carlo simulations using 
the improved two-component $\phi^4$ model on the lattice. It turns out 
that there is 
a discrepancy in the position $x_{min}$ of the minimum of $\theta(x)$:
$x_{min} = -5.3(1)$  \cite{Hu07} and  $x_{min} = -5.43(2)$ 
\cite{VaGaMaDi08}  have to be compared with our result
$x_{min}=- 4.95(3)$ \cite{myCasimir}.

The purpose of the present work is twofold: First we like to figure out 
the performance of the method and secondly we like to check and to 
complement the results of  \cite{myCasimir}. In particular:

We have accurately computed the finite size scaling function of 
the thermodynamic Casimir force $\theta(0)$ at the critical point 
of the three-dimensional bulk system. Our result is consistent with 
the experimental result for $^4$He films \cite{GaCh99} and previous Monte Carlo
simulations \cite{Hu07,MoNi87}.  On the other hand there is a clear discrepancy
with the $\epsilon$-expansion \cite{KrDi92}.

We have demonstrated that the method even allows to compute the 
free energy density of the bulk system. However it seems to be  more 
efficient in this case to integrate the energy density~(\ref{integrateF}).

We have not worked out theoretically how fast $\log z$ converges to 
$\lim_{L_1,L_2 \rightarrow \infty} [f(L_0+1/2,t) - f(L_0-1/2,t)]$. A 
natural guess is that the convergence is exponentially fast in $L_1,L_2$
in the high temperature phase of the film, while in the low temperature 
phase it follows a power law.  For the thickness $L_0=8.5$ we have simulated
at four values of $\beta$ for a large range of $L_1=L_2$ up to $L_1=L_2=1000$.
The results show that the convergence with
$L_1,L_2 \rightarrow \infty$ is no problem in practice.
Our final results for $-\Delta f_{ex}$ at these four values of $\beta$ are
consistent but more accurate than those obtained in \cite{myCasimir}.

Finally we have computed $\theta_{min}$ for several thicknesses, 
where we have taken the values of $x_{min}$ from \cite{myCasimir}. 
This allowed us to improve the estimate $\theta_{min}=1.31(3)$ \cite{myCasimir}
to $\theta_{min}=1.31(2)$. This part of the study nicely shows that the 
virtues of the two method are complementary.

We have not worked out theoretically how the numerical effort increases for a 
given precision with increasing thickness of the film. 
We also have not optimised 
the parameters of the algorithm. However it is quite clear from the simulations
presented here that the method, using our ad hoc choice of the parameters, is 
competitive with previous proposals.

Here we have tested the method at the example of the XY universality class. 
The application to other universality classes, like the Ising or Heisenberg 
universality class is straight forward. On the other hand, the method seems to 
be restricted to films with Dirichlet boundary conditions.

\section{Acknowledgements}
This work was supported by the DFG under the grant No HA 3150/2-1.

\end{document}